\documentclass[runningheads]{llncs}
\usepackage{graphicx}
\usepackage{amssymb}
\usepackage{amsmath}
\usepackage{booktabs}
\usepackage{multirow}
\usepackage{marvosym}
\begin{document}
\title{Brain Network Analysis Based on Fine-tuned Self-supervised Model for Brain Disease Diagnosis}

\author{Anonymous}

\author{Yifei Tang\inst{1,2}, Hongjie Jiang\inst{1,2}, Changhong Jing\inst{2}, Hieu Pham\inst{3} and\\ Shuqiang Wang\inst{2}\textsuperscript{(\Letter)}}

\authorrunning{Y. Tang et al.}

\institute{Southern University of Science and Technology, Shenzhen 518000, China\\ \and
	Shenzhen Institutes of Advanced Technology, Chinese Academy of Sciences, Shenzhen 518055, China\\ \and  College of Engineering and Computer Science and the VinUni-Illinois Smart Health Center, VinUniversity, Hanoi 100000, Vietnam\\
	\email{yf.xie2@siat.ac.cn},
	\email{12333472@mail.sustech.edu.cn},
	\email{ch.jing@siat.ac.cn},
	\email{hieu.ph@vinuni.edu.vn},
	\email{sq.wang@siat.ac.cn}}

\maketitle              
\begin{abstract}
Functional brain network analysis has become an indispensable tool for brain disease analysis. It is profoundly impacted by deep learning methods, which can characterize complex connections between ROIs. However, the research on foundation models of brain network is limited and constrained to a single dimension, which restricts their extensive application in neuroscience. In this study, we propose a fine-tuned brain network model for brain disease diagnosis. It expands brain region representations across multiple dimensions based on the original brain network model, thereby enhancing its generalizability. Our model consists of two key modules: (1)an adapter module that expands brain region features across different dimensions. (2)a fine-tuned foundation brain network model, based on self-supervised learning and pre-trained on fMRI data from thousands of participants. Specifically, its transformer block is able to effectively extract brain region features and compute the inter-region associations. Moreover, we derive a compact latent representation of the brain network for brain disease diagnosis. Our downstream experiments in this study demonstrate that the proposed model achieves superior performance in brain disease diagnosis, which potentially offers a promising approach in brain network analysis research.
\end{abstract}
\section{Introduction}

Functional magnetic resonance imaging (fMRI) is one of the most important imaging technologies in neuroscience \cite{whatcanwedofmri}. The principle of fMRI is based on measuring changes in the blood-oxygen-level-dependent (BOLD) signal induced by neuronal activity using magnetic resonance technology \cite{heeger2002does,logothetis2001neurophysiological}. This groundbreaking technique has led to the fast development of functional brain networks. By calculating the correlation of BOLD signals between different regions of interest (ROIs) in the brain, functional brain networks reveal information about brain activity, particularly in identifying abnormal connectivity \cite{FINGELKURTS2005827}. Since its introduction, it has played a crucial role in brain disease diagnosis \cite{VANDENHEUVEL2010519}. It has greatly contributed to the discovery of biomarkers for brain diseases like Alzheimer's disease, depression, et al, aiding in their diagnosis and treatment, while also supporting research in cognitive neuroscience \cite{greicius2004default,rombouts2005altered,greicius2007resting,chen2024ig}.

In recent years, with the rapid development of artificial intelligence and deep learning, using artificial intelligence to analyze medical images has become a popular research direction\cite{guo2025underwater}. These advanced techniques include convolutional neural networks (CNN) \cite{kawahara2017brainnetcnn,huang2023sd,wang2020ensemble}, graph neural networks (GNN) \cite{li2021braingnn,wein2021graph}, Transformer \cite{kan2022brain,yang2024brainmass} and generative AI \cite{wang2024enhanced}. These computer-aided brain imaging decoding methods have achieved remarkable success. By leveraging multiple nonlinear embeddings and neural network computations, deep learning techniques can model complex inter-regional relationships in the brain and uncover potential dynamic connectivity patterns\cite{liu2010theoretical}.
Despite significant progress in brain network research, these methods require large-scale labeled data and are often trained on specific datasets, which limits their generalizability \cite{wen2023graph}. Additionally, manually annotating a medical dataset is a labor-intensive and time-consuming process, further constraining the potential ability of these methods.
One of the solutions is generative AI. With the widespread application of data augmentation techniques in the field of image processing \cite{chen2024uwformer}, data augmentation by using generative AI effectively generates high-quality and a large amount of medical data \cite{yao2023conditional,li2023generative,yao2025catd}.
Another solution is self-supervised learning. Self-supervised learning paradigms have been proven to be very successful in natural language processing and computer vision \cite{gui2024survey}. It uses auxiliary tasks to extract supervised information from unlabeled large-scale data. By constructing a supervised learning process from unsupervised data, these models are trained to learn general representations used for downstream tasks.

However, a limitation of current methods is their tendency to focus on features from a single dimension, neglecting other dimensions. Brain network computations that concentrate on a single region's features may suffer from reduced applicability and generalizability, limiting their ability to capture a comprehensive representation of brain function. To address this limitation, this study proposes an Adapter module that extends brain region features based on existing foundational brain network models. This enhancement not only improves the generalization capability of the model but also achieves outstanding performance in downstream brain disease diagnosis tasks.

\section{Related Work}
Over the past decade, brain image computing has gained fast development \cite{you2022brain}. In particular, deep learning-based methods and generative AI have made significant progress\cite{wang2020image}. Generative AI generates many annotated medical datasets for training and analysis, which empower medical diagnosis \cite{zuo2024bdht}. Generative Adversarial Network(GAN) \cite{goodfellow2014generative} was first introduced to our eyes. Subsequently, improved GANs are widely used to analyze brain diseases. Kong et al use adversarial learning to analyze the mild cognitive impairment \cite{kong2022adversarial}. Jing et al utilize graph GAN to estimate addiction-related brain connectivity \cite{jing2024estimating}. Diffusion models \cite{ho2020denoising}, which are seen as an advanced generative model, have made a huge success in generating images. Jing et al identify an addicted-relation brain network using a graph diffusion model \cite{jing2024addiction}.

The causes of brain diseases and the mechanisms of brain addiction have always been unknown to us. However, with the introduction of deep learning methods, significant progress has been made in the exploration of brain diseases and brain addiction research \cite{wang2024visualization,gong2023addictive}. Through deep learning methods, researchers study the dynamic changes in brain networks under different brain diseases, providing valuable insights for diagnosis and treatment. Recent advanced deep learning methods are helping researchers identify, analyze, and classify brain images. The first deep learning models applied in this field were Convolutional Neural Networks (CNN), which have been proven to be effective in end-to-end disease diagnosis and have been widely used in brain network analysis. For example, BrainNetCNN \cite{kawahara2017brainnetcnn} introduced novel edge-to-edge, edge-to-node, and node-to-graph convolutional layers to predict cognitive and developmental outcomes in preterm infants based on structural brain networks, SD-CNN \cite{huang2023sd} jointly learned static and dynamic functional connectivity, enhancing brain disease recognition capabilities. Following CNN, Graph Neural Networks (GNN) \cite{scarselli2008graph} gained widespread attention in brain network analysis due to their ability to model complex brain network data. By treating brain regions as nodes and connections as edges, GNN learns abnormal connections in brain networks. These features can not only serve as potential biomarkers but also help to explore the underlying mechanisms of brain diseases. BrainGNN \cite{li2021braingnn} utilized pooling to dynamically learn the importance of different brain regions. Hi-GCN \cite{jiang2020hi} introduced a hierarchical training strategy to improve the representation capacity of Graph Convolutional Networks.

More recently, the Transformer framework \cite{vaswani2017attention} has emerged as the most widely adopted architecture in a variety of machine learning tasks due to its superior representation learning capability. In the field of brain network analysis, the Transformer architecture demonstrates a particularly strong potential, primarily owing to the multi-head self-attention mechanism that enables the model to dynamically capture pairwise relationships among regions of interest (ROIs). Unlike traditional convolutional or recurrent models, which are often limited by local connectivity and sequential dependencies, Transformer-based models can model long-range dependencies and global patterns across the entire brain network. Brain Network Transformer \cite{kan2022brain} analyzed the brain network by using the Transformer framework. BrainMass \cite{yang2024brainmass} established and open-sourced the first foundational model for brain network analysis, marking a breakthrough in the field of brain network analysis.
However, these models always focus on a single dimension of regions of interest(ROIs), ignoring information from other dimensions. Given that the pre-training approach is proven to be effective \cite{dong2024multi}, we propose a fine-tuned self-supervised pre-trained model for brain disease diagnosis.

\section{Method}
\subsection{Overview}
\begin{figure}
	\includegraphics[width=\textwidth]{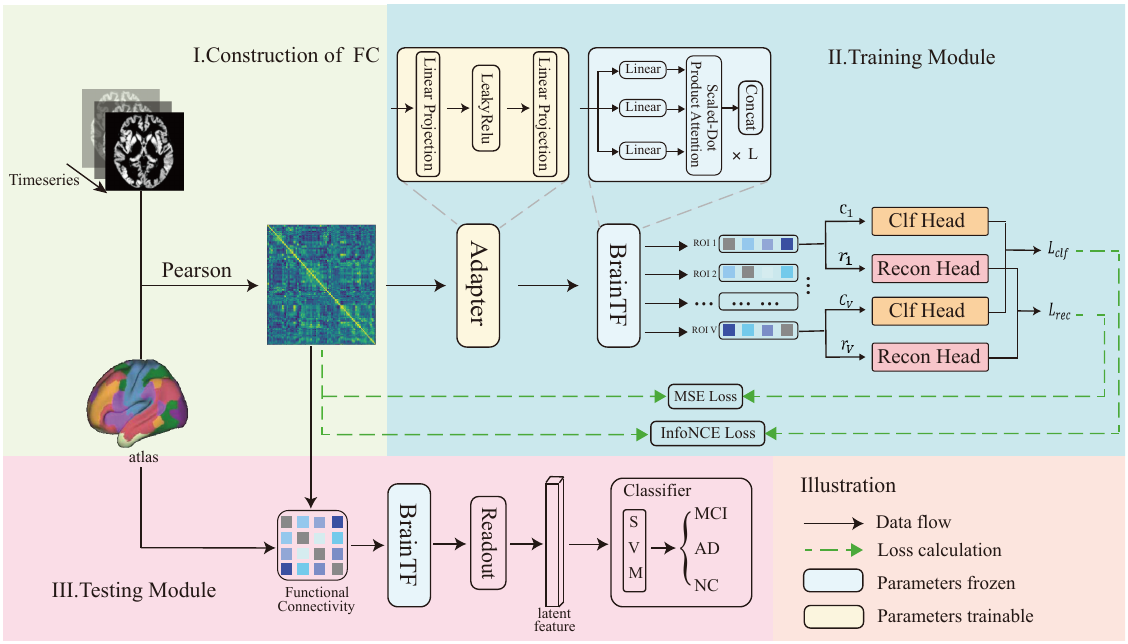}
	\caption{The framework of our method is divided into three parts. I.Construction of FC (Functional Connectivity). II.Training module which composed of Adapter and Brain network Transformer encoder. III.Testing module is for downstream tasks. The dark arrow means the overall data flow. Green dashed line indicates loss calculation.} \label{fig1}
\end{figure}

We propose an Adapter module to expand the dimension of brain network inputs. The overview of our method has shown in Fig.\ref{fig1}. We first get BOLD signal $X$ by mapping the fMRI onto a certain template with $V$ regions of interest. We apply Pearson Coefficient Correlation to measure the functional brain network $R^{V \times V}$. The Adapter is composed of two linear projections which is used to expand the dimension of brain network. BrainTF is a pre-trained Transformer encoder. The data flow is shown in Fig.\ref{fig1}, we calculate two losses for the backward propagation. During the training, we freeze the parameters of BrainTF and train the Adapter module. As for downstream tasks, we aim to get the classification result $y$, which represents the predicted type. We freeze the BrainTF encoder and use readout function to extract latent feature. The latent feature is used to fed the Support Vector Machine(SVM) classifier to do the downstream tasks.

\subsection{Adapter}
Our Adapter use a feed-forward network with two hidden layers. We apply two linear projections. Assume the input is $R^{V \times V}$, the goal of Adapter is build the mapping function $f:R^{V \times V}\rightarrow R^{V \times B}$, where B represents the output dimension.

\subsection{Brain network Transformer encoder}
Transformer-based models led to a huge success in fields including computer vision, natural language processing \cite{kalyan2021ammus,khan2022transformers}. Here, we will simply introduce the transformer structure.

In this study, the input brain network $X \in R^{V \times V}$ is seen as a sequence ($x_1,...,x_V$). Multi-head self attention is applied to calculate the inter-ROIs relations. As for the whole transformer encoder, the number of heads is $H$, the depth of Transformer block is $D$. The equation \ref{eq1} shows the calculation process:
\begin{equation}
	O = MHSA(X) \in R^{V \times V} \label{eq1}
\end{equation}
where O is output. For each layer $d$, we first calculate query $Q^{d,h}$, key $K^{d,h}$ and value $V^{d,h}$ where $h$ indicates $h$-th head:
\begin{equation}
	Q^{d,h} = O^{h-1}W^{d,h}_q \label{eq2}
\end{equation}
\begin{equation}
	K^{d,h} = O^{h-1}W^{d,h}_q \label{eq3}
\end{equation}
\begin{equation}
	V^{d,h} = O^{h-1}W^{d,h}_q \label{eq4}
\end{equation}
For each head, the output is calculated as:
\begin{equation}
	O^{d,h} = Softmax(\frac{Q^{d,h} \times (K^{d,h})^T}{\sqrt{d_k}})V^{d,h} \label{eq5}
\end{equation}
where softmax function is used to convert vector values into probabilities. The formula of Softmax function is shown below:
\begin{equation}
	\sigma(z_i) = \frac{exp(z_i)}{\sum_{j=1}^{V}exp(z_j)} \label{eq6}
\end{equation}
Finally, we get the output by integrating each heads' output:
\begin{equation}
	O = concat(O^{D,1},O^{D,2},...,O^{D,H}) \label{eq7}
\end{equation}
In this study, the BrainTF we used is pre-trained under thousands of subjects which contain healthy controls and illness. Unlike traditional pretrained models, foundation models trained on large-scale datasets can perform a diverse range of tasks using a single set of model weights \cite{zhang2024challenges}.

\subsection{Optimization}
We apply two losses in the training phase. There are two tasks we aim to do: reconstruct the brain network and do the disease diagnosis. So We calculate the classification head and reconstruct head respectively to predict the features. To employ this, we implement mean square error loss for reconstruction head and InfoNCE loss \cite{oord2018representation} for classification head. Mean square error loss measure the difference between the input and output brain network. InfoNCE loss makes the output representation closer to the positive sample, which is beneficial for classification tasks. For InfoNCE loss, we aim to find the mapping ($q$,$k^+$) that positive to classification. The equation of two losses are shown below:
\begin{equation}
	L_r = \frac{1}{N}\sum_{i = 1}^{N}(y_i - \hat{y_i})^2 \label{eq8}
\end{equation}
\begin{equation}
	L_c = -log\frac{exp(sim(q,k^+)/\tau)}{\sum_{j = 1}^Nexp(sim(q,k^+)/\tau)} \label{eq9}
\end{equation}
where $sim$ means calculate the cosine similarity and $\tau$ means temperature coefficient.

For backward propagation,we employ a weighted loss function, adding the balance parameters $\lambda_c$ and $\lambda_r$ during the training:
\begin{equation}
	L = \lambda_cL_c + \lambda_rL_r \label{eq10}
\end{equation}
This formula enables model to learn the dependencies across brain connectome patterns and enhance the performance for downstream tasks.

\section{Experiments}
\subsection{Datasets and Experiment Design}
\subsubsection{Datasets.}
In this experiment, we use the Alzheimer’s Disease Neuroimaging Initiative(ADNI) datasets \cite{mueller2005alzheimer}, which contains 64 Alzheimer's disease(AD), 135 mild cognitive impairment(MCI) group and 263 normal controls(NC).
All fMRI data were mapping by the AAL atlas \cite{tzourio2002automated} parcellations into 90 ROIs under anatomical space(MNI152). Before experiment, all data was preprocessed to obtain the same time fMRI signal.
\subsubsection{Implement Details.}
We train our model on a single NVIDIA Geforce RTX 3080. The PyTorch backend is implemented to conduct the experiment. The optimizer is set to Adam \cite{kingma2014adam} with learning rate set to $3 \times 10^{-4}$ and weight decay set to ${5 \times 10^{-5}}$. For optimization, we set $\lambda_c$ and $\lambda_r$ to 0.2 and 5 in Eq.~\ref{eq10}. The hidden size of our Adapter is 1024. The epoch of training is set to 500. In the downstream tasks, we apply the SVM classifier for prediction. To ensure the robustness of our results, we repeat the experiment three times and take the average.
\subsubsection{Metrics}
We assess the performance of our model using accuracy, sensitivity, specificity, and F1-score. In addition, we show the reconstruct brain network compared to origin brain network, allowing us easy to assess the results.

\subsection{Evaluation}
\begin{figure}
	\centering
	\includegraphics[width=0.8\textwidth,height=16cm]{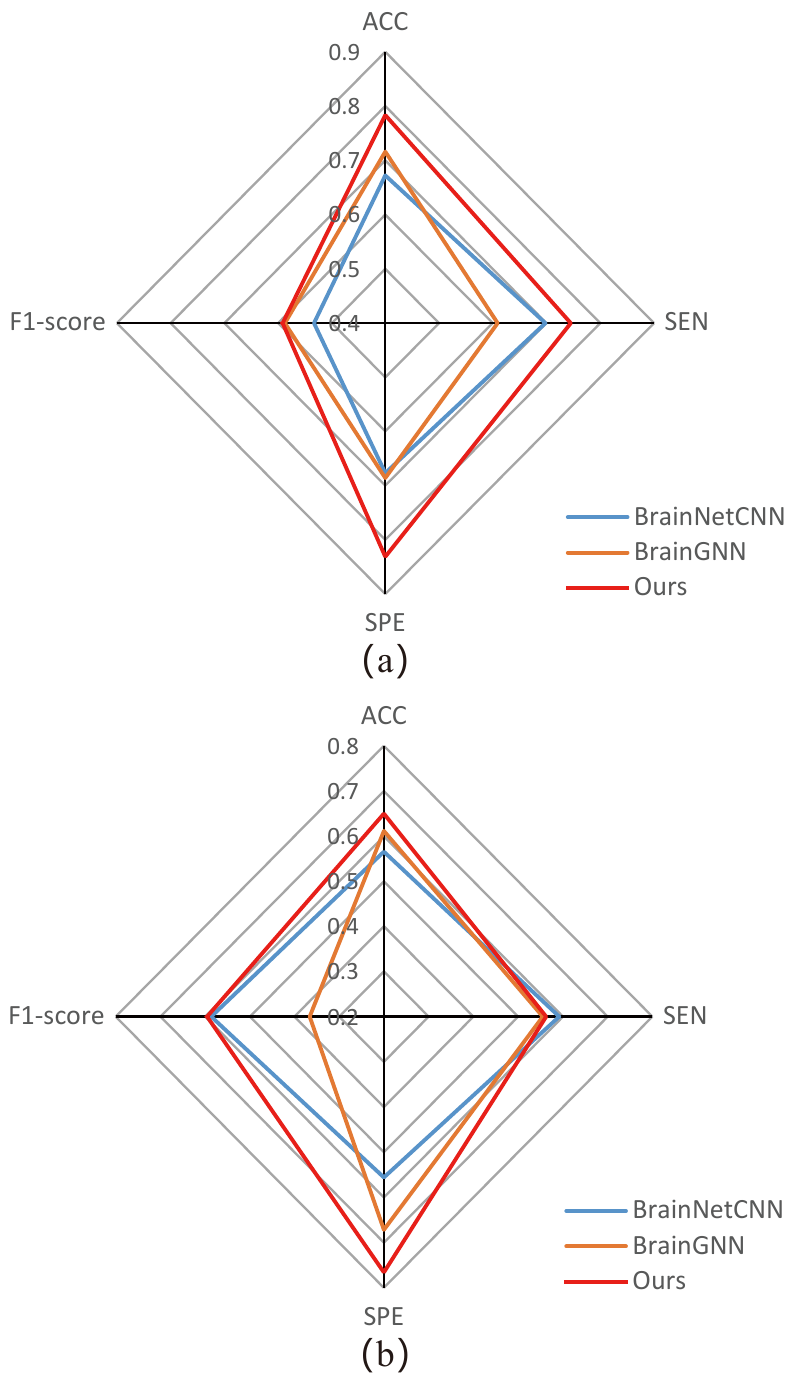}
	\caption{Radar plot about our evaluation of two baseline models. (a) shows the metrics in classification task AD vs NC. It is obvious that our model perform best in four metrics. (b)shows the metrics in classification task MCI vs NC. Our model perform best in acc, spe and F1-score.} \label{fig2}
\end{figure}

To evaluate the effectiveness of our model, we compared our model to two other baseline models: BrainNetCNN and BrainGNN. Table~\ref{tab1} and Table~\ref{tab2} show the classification performance with AD vs NC, MCI vs NC respectively. And to better visualize the results of our experiment, we employed visualization techniques. Fig.~\ref{fig2} contains two radar plots, demonstrating our results more concisely and clearly.

\begin{table}
	\caption{Classification results of our experiment between AD and NC.}
	\centering
	\setlength{\tabcolsep}{5pt}
	\begin{tabular}{l|c|c|c|c}
		\hline
		Task & \multicolumn{4}{c}{AD vs NC}\\
		\hline
		Metric & ACC & SEN & SPE & F1-score \\
		\hline
		BrainNetCNN & 67.21\% & 69.76\% & 67.69\% & 0.5333 \\
		\hline
		BrainGNN & 71.63\% & 60.92\% & 68.57\% & 0.5872\\
		\hline
		Ours & \textbf{78.35\%} & \textbf{74.55\%} & \textbf{83.12\%} & \textbf{0.5912}\\
		\hline
	\end{tabular}
	\label{tab1}
	\vspace{10pt}
	\caption{Classification results of our experiment between MCI and NC.}
	\centering
	\setlength{\tabcolsep}{5pt}
	\begin{tabular}{l|c|c|c|c}
		\hline
		Task & \multicolumn{4}{c}{MCI vs NC}\\
		\hline
		Model & ACC & SEN & SPE & F1-score \\
		\hline
		BrainNetCNN & 56.51\% & \textbf{59.12\%} & 55.61\% & 0.5867 \\
		\hline
		BrainGNN & 61.16\% & 55.46\% & 67.14\% & 0.3662 \\
		\hline
		Ours & \textbf{64.96\%} & 56.21\% & \textbf{76.62\%} & \textbf{0.5952}\\
		\hline
	\end{tabular}
	\label{tab2}
\end{table}

In the classification task between AD and NC, our model achieved approximately 11\% and 7\% higher accuracy compared to BrainNetCNN and BrainGNN, respectively. In the classification task between MCI and NC, our model achieved 8\% and 3\% higher accuracy. Besides, our model also performs better in other metrics. However, the accuracy of MCI and NC classification task is relatively low. On the one hand, we believe that the differences in brain region connectivity between individuals with mild cognitive impairment and healthy controls are less pronounced compared to those between Alzheimer's disease patients and healthy controls. On the other hand, our dataset consists of fMRI data from a single scan, which may affect our model's ability to learn the characteristics of brain regions. In summary, by utilizing Adapter to expand the dimension of BrainTF, our model achieves better performance.

\subsection{Brain Network Analysis}
\begin{figure}
	\centering
	\includegraphics[width=\linewidth]{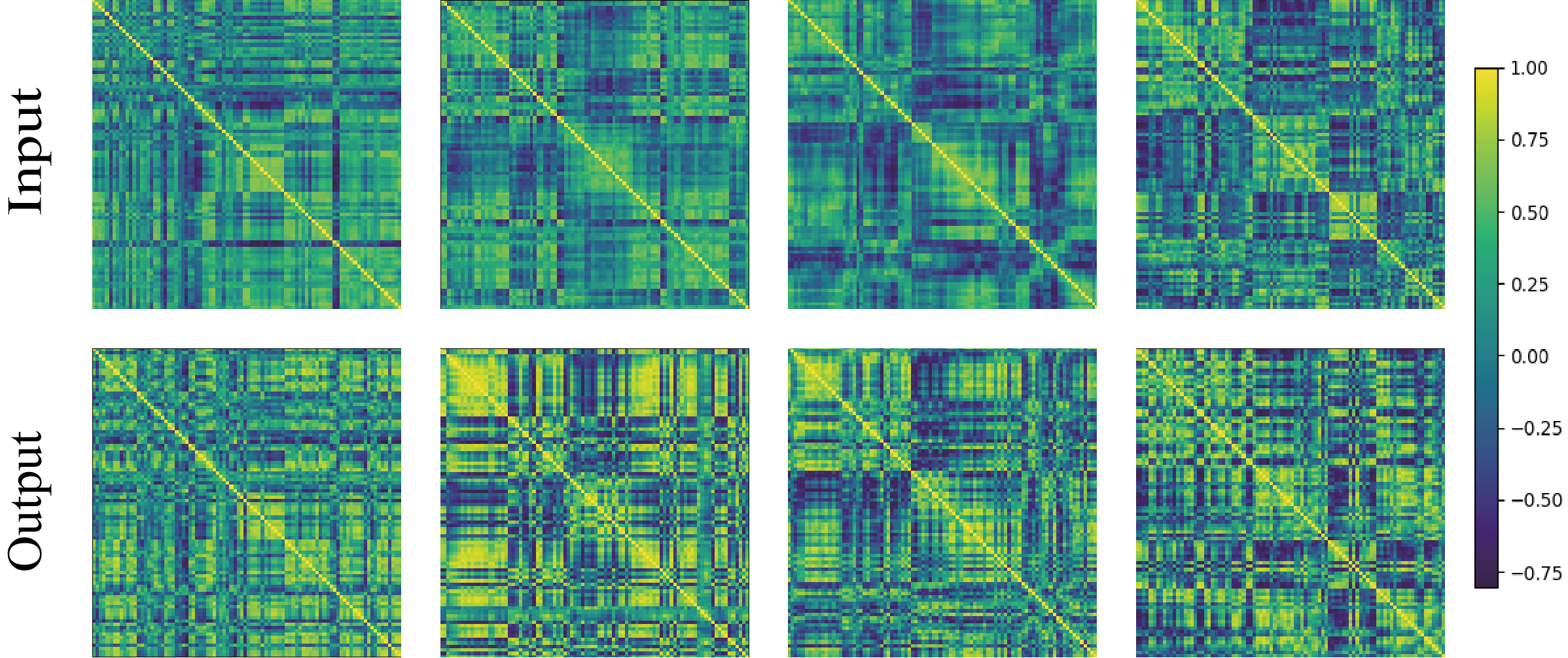}
	\caption{Input and Output brain network from AD patient. The brain network constructed by our model can effectively capture the inter-regional relationships of the brain in Alzheimer's disease patients, providing strong evidence to support our performance in downstream tasks.}
	\label{fig3}
\end{figure}

We not only focus on the metrics evaluation, but also observe the brain network. We presented several inputs and outputs of AD in Fig.~\ref{fig3}, for example. In brain regions with relatively high correlation, our model can successfully capture the connectivity between regions. However, in some less active brain region connections, our model also outputs high correlation. This may be due to differences in brain activity among different subjects, leading to variations in the model’s output. Overall, our model is effective in capturing the characteristic brain region features of Alzheimer's disease, which strongly support our model's capability of brain disease diagnosis.

\subsection{Ablation Study}
In Eq.~\ref{eq10}, we apply two parameters to calculate loss. In ablation study, we aim to discover whether parameters $\lambda_c$ and $\lambda_r$ necessary to our model. The ablation studies are summarized in Table.~\ref{tab3}. We can see that in the classification task between MCI and NC, $\lambda_c$ has a considerable impact on accuracy, but little impact on AD and NC classification task. Considering there are two tasks we aim to accomplish: build brain network and brain disease diagnosis. There may be an issue of imbalanced backpropagation, as the magnitude of MSE and InfoNCE loss in our loss function differs by approximately a factor of 10. Without any of it, the performance of our model degrade. And when we combined two parameters, the performance are further improved.
\begin{table}
	\caption{Ablation studies on the elements of our model with the performance of accuracy.}
	\centering
	\setlength{\tabcolsep}{5pt}
	\begin{tabular}{c|c|c|c}
		\hline
		\multicolumn{2}{c|}{Task} & AD vs NC & MCI vs NC \\
		\hline
		$\lambda_c$ & $\lambda_r$ & ACC & ACC \\
		\hline
		& $\checkmark$ & 77.32\% & 59.83\% \\
		\hline
		$\checkmark$ & & 75.26\% & 61.54\% \\
		\hline
		$\checkmark$ & $\checkmark$ & \textbf{78.35\%} & \textbf{64.96\%} \\
		\hline
	\end{tabular}
	\label{tab3}
\end{table}

\section{Conclusion}
In this paper, we propose a new model to explore the brain region characteristics of Alzheimer's disease patients and apply them to brain disease diagnosis. We introduce an Adapter that expands the dimensionality of brain region features and demonstrates excellent generalization capabilities. We conducted a detailed discussion on important aspects such as the Adapter and the encoder. Our experimental results outperform the comparison methods, indicating that our model has an advantage in feature extraction. More importantly, our research may contribute to the exploration of brain networks in the field of neuroscience. In the future, we will continue to investigate brain networks.

\subsubsection{Acknowledgment.}
This work was supported by the National Natural Science Foundations of China under Grant 12326614 and 62172403, and the Distinguished Young Scholars Fund of Guangdong under Grant 2021B1515020019.

\bibliographystyle{splncs04} 

\end{document}